\documentclass[prl,aps,twocolumn,showpacs,superscriptaddress,longbibliography]{revtex4}
\usepackage{graphicx}
\usepackage{color}
\definecolor{darkblue}{rgb}{0,0.02,0.45}
\usepackage[colorlinks=true,citecolor=darkblue]{hyperref}

\begin{document}

\title{Pseudospin-lattice coupling and electric control of the square-lattice iridate $\rm Sr_2IrO_4$}

\author{Feng Ye}
\author{Christina Hoffmann}
\author{Wei Tian}
\affiliation{Neutron Scattering Division, Oak Ridge National Laboratory, Oak Ridge, Tennessee 37831, USA}
\author{Hengdi~Zhao}
\author{G.~Cao}
\affiliation{Department of Physics, University of Colorado at Boulder, Boulder, Colorado 80309, USA}
\date{\today}

\begin{abstract}
    $\rm Sr_2IrO_4$ is an archetypal spin-orbit-coupled Mott insulator and has
    been extensively studied in part because of a wide range of predicted
    novel states. Limited experimental characterization of these states thus
    far brings to light the extraordinary susceptibility of the physical
    properties to the lattice, particularly,  the Ir-O-Ir bond angle.  Here,
    we report a newly observed microscopic rotation of the IrO$_6$ octahedra
    below 50~K measured by single crystal neutron diffraction. This sharp
    lattice anomaly provides keys to understanding the anomalous
    low-temperature physics and a direct confirmation of a crucial role that
    the Ir-O-Ir bond angle plays in determining the ground state. Indeed, as
    also demonstrated in this study, applied electric current readily weakens
    the antiferromagnetic order via the straightening of the Ir-O-Ir bond
    angle, highlighting that even slight change in the local structure can
    disproportionately affect the physical properties in the
    spin-orbit-coupled system.
\end{abstract}
\maketitle

Strong spin-orbit interactions (SOI), along with appreciable Coulomb
interactions, crystalline electric field, and large orbital hybridization in
5$d$-electron based oxides has produced a wide range of quantum phenomena such
as spin liquid phases \cite{okamoto07}, superconductivity
\cite{wang11c,watanabe13}, and Kitaev magnetism
\cite{jackeli09,price12,singh12}. One prominent example is the observation of
the so-called effective $j_{\rm eff}=$1/2 Mott insulating state in iridates
\cite{kim08,moon08,kim09}, where the magnetism is attributed to an isotropic
pseudospin \cite{rau16, winter17, hermanns18}.  Unlike the situation in the
$3d$ transition metal oxides which have a distinct energy scale in both
orbital and spin parts and sequential orbital and magnetic orders, the notion
of spin-orbital separation in the $5d$ systems is no longer valid due to the
strong SOI. Furthermore, the Jahn-Teller effect, which measures the couples
between the lattice and orbital degrees of freedom and is a common occurrence
in $3d$-electron based oxides, remains largely unexplored in the iridates.
However, recent experimental and theoretical studies indicate that the
magnetic properties and low-energy spin dynamics \cite{porras19,liu19}, can be
better explained using a pseudospin-lattice coupling mechanism. The
introduction of such a term in the spin Hamiltonian not
only correctly describes the metamagnetic transitions \cite{porras19}, but
also explains a finite in-plane magnon gap \cite{gim16,gretarsson17,calder18}.
It also predicts an orthorhombic structural distortion in the order of
10$^{-4}$, which can be probed using a high resolution Larmor precession
technique \cite{li17c}.

The extraordinarily strong coupling between the electronic and structural
properties in the iridates provides unprecedented opportunities to uncover
novel quantum phases by merely modifying the local structure. It is
particularly interesting that application of hydrostatic pressure steadily
reduces the charge gap of approximately 500~meV and diminishes the
antiferromagnetic (AFM) state existent in the square lattice $\rm Sr_2IrO_4$
(Sr-214), yet retains the insulating state \cite{zocco14,chen20,samanta20}.  A
possible quantum paramagnetic state at pressures greater than 20 GPa is
proposed, attributing to the suppression of the inter-layer exchange coupling
and enhanced magnetic frustration within the IrO$_2$ layer \cite{haskel20}.
Alternative approaches to modify the structure using strain engineering are
also reported in epitaxial grown thin films of Sr-214
\cite{lupascu14,miao14,lu15a,kim16a,seo19}, where a compressive or tensile
strain can drastically enhance or reduce the transition temperature $\rm
T_N$. In some cases, the strengthened in-plane exchange interaction can
promote a short range magnetic correlation well above the nominal transition
\cite{seo19}. In this paper, we report a newly observed lattice anomaly at
50~K. Given the exceptionally strong coupling between the lattice and physical
properties in Sr-214, this structural anomaly, which has escaped all previous
studies until now. Our study provides much-needed keys to better understand the
low-temperature magnetic and transport behavior, which remains a focus of
current studies on the iridates. Indeed, the susceptibility of the physical
properties to the lattice is further revealed by the application of in-plane
$dc$ electric current as a new stimulus that straightens the Ir-O-Ir bond angle,
thus suppresses the canted AFM order. These results reveal
crucial insights into the crystal structure and demonstrates that the subtle
changes in the crystal structure, whether ambient or induced, dictate the physical
properties in this spin-orbit-coupled Mott insulator.

Single crystals of Sr-214 were grown using the self-flux method \cite{cao98,
cao18a}.  The magnetic susceptibility and specific heat were measured using a
Quantum Design Magnetic Property Measurement System. The neutron diffraction
measurements were carried out at the HB1A triple axis spectrometer at the High
Flux Isotope Reactor and the TOPAZ diffractometer at the Spallation Neutron
Source, ORNL. For measurements at HB1A, a sample assembly of over 40
single crystals (mass $\sim$ 25~mg, mosaicity $\sim1.5^\circ$) was aligned at
various scattering planes to probe the magnetic and nuclear reflections. A
closed-cycle refrigerator was used to regulate the sample temperature ($T$).
For measurements at TOPAZ, a plate-like single crystal with dimensions of
$\rm 1.8~mm\times1.8~mm\times0.55~mm$~was used. The sample temperature is
controlled using the Oxford cryostream cooler Cobra 800. Electric current
was applied using the power supply from BioLogic science potentiostat with a
voltage range of $\pm 12$~V and current range $\pm 200$~mA.

Figures~1(a)-1(b) show the main structural features of the Sr-214 with Ir ions
forming a square-lattice network.  The system exhibits a characteristic
staggered rotation of $\rm IrO_6$ octahedra along the $c$ axis at about $\approx
11.8^\circ$ \cite{huang94,ye13}.  Because of the layered structure, the
corresponding $\rm IrO_6$ octahedra elongate slightly along the $c$-axis
causing a crystal electric field (CEF) level $\Delta_{\rm CEF}\sim200$ meV
\cite{moon06,jackeli09}. Neutron scattering studies reveal a long-range
magnetic order below 240~K [Fig.~1(e)]. Magnetic structure refinement
indicates that the spins of Ir ions form a canted AFM configuration.
The in-plane magnetic components closely track the rotation of the $\rm IrO_6$
\cite{ye13,boseggia13}, which results directly from the SOI; the canting
angle is governed by a characteristic ratio, $\phi\sim D/2J$, of the
antisymmetric Dzyaloshinskii-Moriya (DM) interaction $D$ to the isotropic
Heisenberg interaction $J$ \cite{jackeli09}.

\begin{figure}[ht!]
 \includegraphics[width=3.3in]{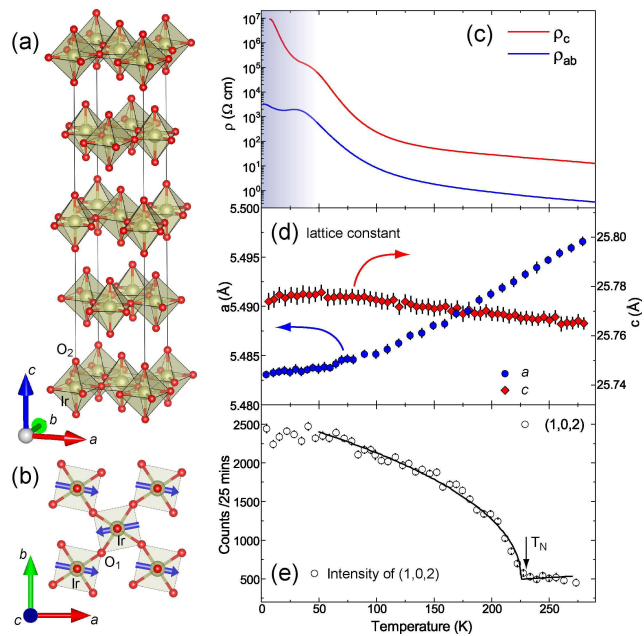}
    \caption{(a) The crystal structure of $\rm Sr_2IrO_4$ in the tetragonal
    setting. For simplicity, only the iridium and oxygen atoms are shown. (b)
    The connecting $\rm IrO_6$ network is viewed along the $c$ axis. The
    $T$-dependence of (c) the in-plane and $c$-axis resistivity
    $\rho_{ab}(T)$ and $\rho_{c}(T)$, (d) the lattice parameters $a$ and $c$
    measured from neutron diffraction, and (e) the strongest AFM 
    peak (1,0,2). The solid line in panel (e) is a guide to the eye.
    }
\end{figure}

It is known that the onset of long range magnetic order at $\rm T_N$ causes no
anomaly in $\rho_{ab}$ or $\rho_c$ \cite{cao98,chikara09}, specific heat (data
not shown), and the lattice parameters [Figs.~1(c)-1(d)]. The conventionally
anticipated correlation between the structural and physical properties is
conspicuously  missing at $\rm T_N$. However, such a correlation is instead established
below a newly observed lattice anomaly, $\rm T_M=50$~K.  As shown in
Fig.~1(c), both the in-plane and $c$-axis resistivity exhibit noticeable kinks
below $\rm T_M$.  Recently, independent experiments have confirmed the anomalous
character at low $T$. Among them, a magnetic study shows that the residual
magnetization increases below $\rm T_N$, peaks at $\rm T_M$, and diminishes on
further cooling \cite{bhatti19}.  Such behavior was initially attributed to
the reorientation of the canted spin moment \cite{chikara09}.  A muon spin
rotation study by Franke {\it et al.} shows that the oscillation of single
precession signal below the magnetic transition splits into high and low
frequency components below 20~K \cite{franke11}.  The authors of
Ref.~[\onlinecite{franke11}] suggest that the modification of the magnetic
structure leads to structurally equivalent muon sites that experience
increasingly distinct local field at low temperatures.

Nevertheless, the abnormal low-$T$ behavior demands a detailed examination of
the crystal structure. The neutron diffraction measurements indicate that the
system remains tetragonal at all temperature and does not show a lower
symmetry below the magnetic transition \cite{huang94,ye13,dhital13,ye15}.
Since the neutron coherent scattering length of the light oxygen atom ($b_{\rm O}
= 5.8$ fm) is comparable to the much heavier strontium ($b_{\rm Sr}=7.02$ fm)
and iridium ( $b_{\rm Ir}=10.6$ fm), the subtle change of the $\rm IrO_6$
rotation can be readily characterized using neutron diffraction and used to
quantify its coupling to the magnetic order.  The structure factor of
the nuclear scattering at a wavevector transfer $\vec{q}$ is
\begin{eqnarray}
    |F(\vec{q})|^2 & = & |\sum_i b_i \exp(\vec{r}_i \cdot \vec{q})|^2,
\end{eqnarray}
where $b_i$ is the coherent scattering length for individual atom inside the
unit cell and $\vec{r}_i$ is the corresponding atomic coordinate.
The rotation of the IrO$_6$ octahedra leads to the nonzero peak
intensities forbidden for the space group $I4/mmm$, e.g.,~$(2,1,2n+1)$,
and is solely determined by the in-plane oxygen atom O$_1$ located at
($1/4-\delta$,$1/2-\delta$,$1/8$) and equivalent positions, where
$\delta\approx0.05$ characterizes the deviation from the undistorted $\rm
IrO_6$ scenario.  The intensity of the superlattice peak (2,1,1), $I_{211}$, can
be derived as $b_O (\sin \delta+\sin 3\delta)^2$ and is proportional to
$\delta^2$ in the small rotation limit.

\begin{figure}[thb!]
     \includegraphics[width=3.0in]{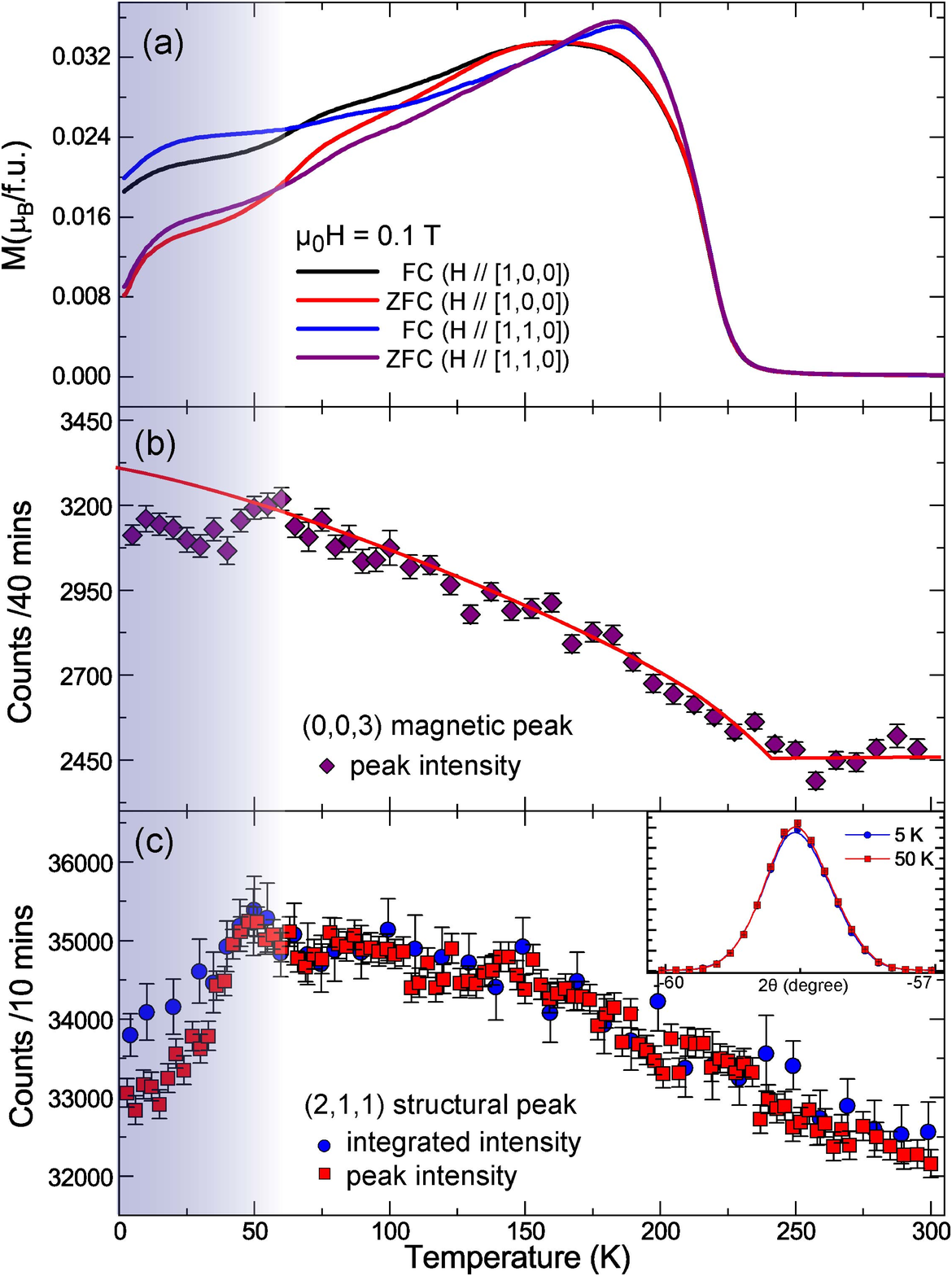}
     \caption{ (a) The magnetization $M$ in the zero-field-cooled (ZFC) and
     field-cooled (FC) protocols, with a magnetic field of $0.1$~Tesla applied
     along the $[1,0,0]$ and $[1,1,0]$ directions. (b) The $T$-dependence of
     the peak intensity of the magnetic (0,0,3) reflection which probes the
     canted spin component. The solid line is a guide to the eye. (c) The
     $T$-evolution of the structural peak (2,1,1) associated with the in-plane
     rotation of the $\rm IrO_6$ octahedra. The (blue) circles are the
     integrated intensities, the (red) squares are the peak intensities. Inset
     shows the comparison of the $\theta-2\theta$ scans at 5~K and 50~K. Both
     peaks are fit using Gaussian profile.
     }
\end{figure}

The red squares in Fig.~2(c) show the $(2,1,1)$ peak intensity 
upon cooling. The gradual increase in intensity implies an
enhancement of the $\rm IrO_6$ rotation. Surprisingly, the monotonic evolution
experiences an abrupt change below 50~K.  To further verify the feature,
$\theta-2\theta$ scans across the peak are performed (blue circles).  The
marked decrease at low-$T$ reveals that the suppression of IrO$_6$ octahedral
rotation is intrinsic.  Comparing to 50~K, the value at 5~K is reduced by 5\%,
which translates to a suppression of rotation angle by 0.3$^\circ$.  As
the moment strictly follows the octahedral rotation, one expects a sudden
reduction of the rotation or straightening of the Ir-O-Ir bond angle. This can
be probed by examining the magnetic reflection $(0,0,3)$ that is
exclusively sensitive to the canted component \cite{ye13}. Fig.~2(b) displays
the $T$-dependence of the magnetic peak with long counting time.  A
similar kink is observed below 50~K, reinforcing the structural anomaly in
Fig.~2(c). This is also consistent with the abrupt reduction of the
magnetization shown in Figure 2(a) with an external magnetic field of $\rm
\mu_0H = 0.1$~T applied along either the $[1,0,0]$ or the $[1,1,0]$
directions, which also agrees well with our earlier report \cite{ge11}.  These
data point out a close association between the crystal structure and magnetic
order, confirming an essential role of the pseudospin-lattice
coupling that dictates the low-energy magnetic properties.

\begin{figure}[thb!]
     \includegraphics[width=3.1in]{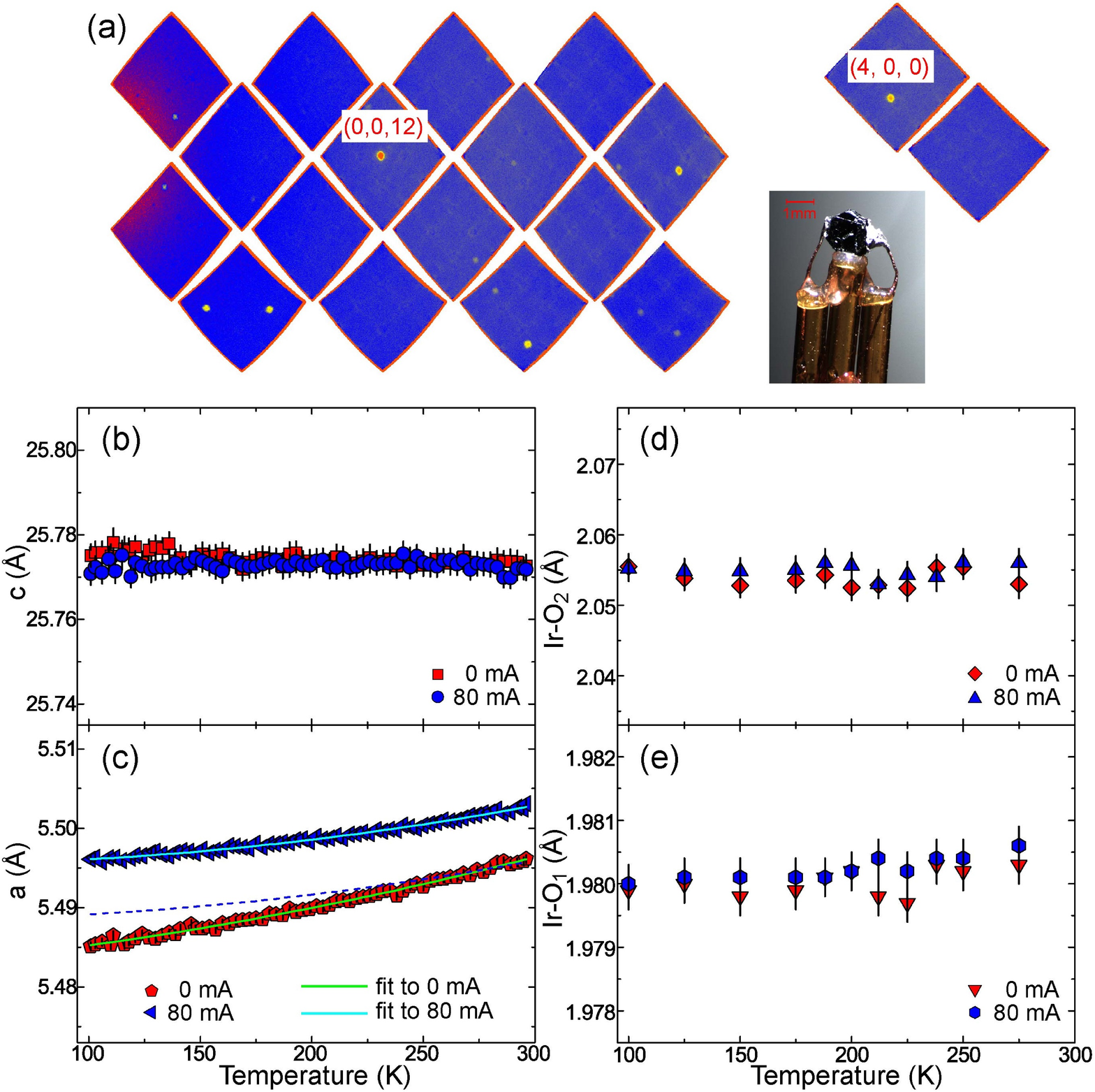}
     \caption{(a) The TOPAZ instrument view of the diffraction experiment. The
     sample is oriented to allow optimal collection of desired reflections
     based on the physical layout of detector modules.  Inset shows the
     experimental setup for applying electric current. The temperature is
     regulated using active cryo-stream cooling to ensure the sample is in
     thermal equilibrium \cite{zhao19a}. The $T$-dependence of the lattice
     parameters $c$ [panel (b)] and $a$ [panel (c)] without and with applied
     electric current of 80~mA.  The solid green and light blue curves are
     quadratic fits of the lattice parameters. The light blue curve is also
     shifted down vertically by $0.007~\rm \AA$ (the dashed blue curve) to
     match the $I=0$~mA data above the magnetic transition.  The corresponding
     refined out-of-plane Ir-O$_2$ and  in-plane Ir-O$_1$ bond distance are
     shown in panels (d)-(e).
     }
\end{figure}

This key characteristic of the lattice degree of freedom provides an effective
``knob'' to control the physical properties of the iridates using small
external stimuli that readily couple to the lattice.  Electrical current, as a
novel stimulus, is particularly effective in controlling the lattice, and thus
the magnetic properties \cite{cao18}. The previous x-ray study has shown a
prominent lattice expansion for the undoped Sr-214 under electric current,
accompanied by remarkable reduction in both transition temperature and
in-plane magnetization. No structural anomaly is observed in Tb-doped Sr-214
where long range magnetic order is absent \cite{wang15}. To better
characterize the structural response, we perform single crystal neutron
diffraction measurements with in situ electric current application.  The
orientation of the Sr-214 crystal is determined using the x-ray Laue method to
ensure the currents are applied in the basal plane [inset of Fig.~3].  Since
the diffractometer TOPAZ has a white incident neutron beam with large
reciprocal coverage, the sample is mounted such that the characteristic
$(4,0,0)$ and $(0,0,12)$ peaks can be simultaneously accessed
[Fig.~3(a)]. In such fixed-orientation configuration, we collected a total
number of around 2000 nuclear reflections at each temperature that allow a
reliable structural determination. A electric current of 80 mA (current
density $\sim$ 8~A/cm$^2$) is employed at 300 K and the sample is cooled to
100~K while the electric current is maintained. The crystal is subsequently
warmed to 300~K at a rate of 0.5~K/min.  Figs.~3(b)-3(c) display the
$T$-dependence for the out-of-plane and in-plane lattice parameters with and
without electric current, respectively. It is evident that the $c$-axis
lattice parameter is insensitive to the current. In contrast, the in-plane
lattice parameter displays a dramatic response to the current, consistent with
the fact that the magnetic moments lie within the basal plane. The overall
$T$-dependence of lattice parameter $a$ at 80~mA stays above the one at $I=0$.
While both data can be well described by a quadratic form, a slope change near
the AFM transition is clearly visible [see the dashed curve in Fig.~3(c)]. The
detailed structure refinement reveals that the Ir-O bond distances remain
unchanged at $I=80$~mA.  The corresponding $\rm IrO_6$ octahedron also remains
slightly elongated, with a ratio of the Ir-O$_2$ to the Ir-O$_1$ bond
distances being 1.04 throughout the temperature range measured.

\begin{figure}[thb!]
     \includegraphics[width=3.2in]{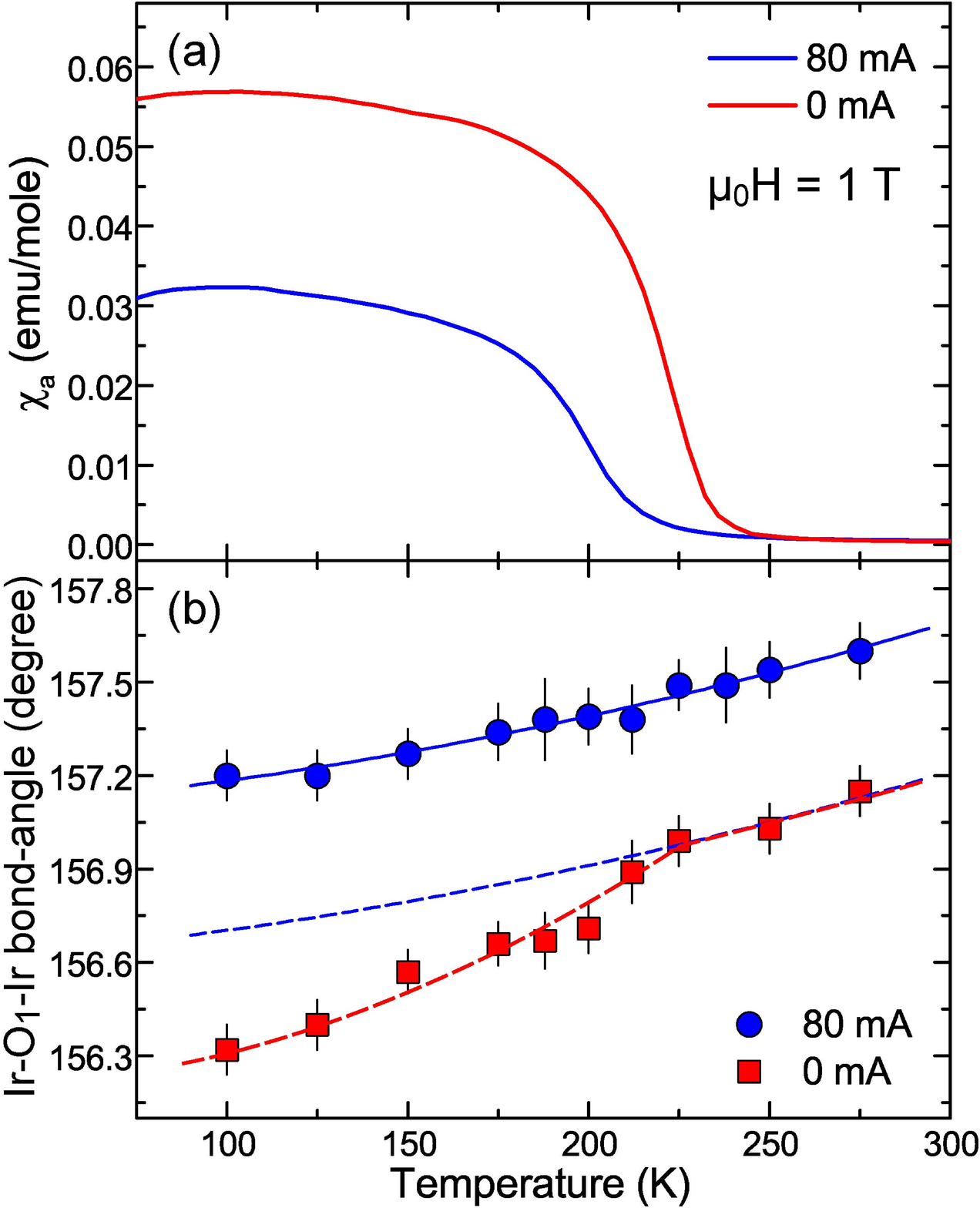}
     \caption{ The magnetic susceptibility $\chi(T)$ with electric current of
     $I$=0 and 80 mA applied in the basal plane. (b) The $T$-dependence of
     refined in-plane Ir-O$_1$-Ir bond angle with and without electric current.
     }
\end{figure}

Although the $\rm IrO_6$ octahedra remain rigid irrespective of the applied
electric current, the in-plane Ir-O-Ir bond angle between the corner-sharing
$\rm IrO_6$ octahedra undergoes a considerable modification. Figure 4(b)
compares the refined bond angle between the two situations. The angle at 0~mA
stays smaller than 157$^\circ$ and decreases with further reduction of
temperature. A slope change occurs at the magnetic transition $\rm T_N$ at
$I=0$~mA.  In sharp contrast, the bond angle becomes notably straighter
or relaxed by 0.3$^\circ$ when a current of $I=80$~mA is applied in the basal
plane. In addition, these bond angle values show a smooth decrease upon
cooling; the anomaly observed near $\rm T_N$ at the ambient condition is
significantly suppressed. Indeed, the corresponding in-plane magnetic
susceptibility displays the behavior consistent with the structural data
[Fig.~4(a)].  The $\chi_a$ at 80~mA shows reduction in amplitude along with
the suppression and broadening of the transition temperature.  The reduced
magnetization is expected because the DM interaction $D$ becomes smaller as
the Ir-O-Ir bond angle relaxes due to the applied current (recall
$\phi\sim D/2J$).

Tuning the physical properties, in particular magnetism using
electric current, is highly desirable. It opens a new frontier for studies of
the correlated and spin-orbit-coupled systems \cite{cao18,nair19,okuyama19}.
For instance, a reorientation of the AFM order is observed in the
metallic $\rm Fe_{1/3}NbS_2$ at a remarkable low current density
($\sim$ 10$^4$~A/cm$^2$) showing the potential to build 
AFM spintronics devices \cite{nair19}. Spatially inhomogeneous
rotation of the skyrmion lattice in the metallic helimagnet MnSi occurs when a
current flows across the sample. This emphasizes the role of friction near
the sample edges for skyrmion-based applications \cite{okuyama19}. Much efforts
have also been devoted to {\it insulating} materials such as ruthenate
oxide $\rm Ca_2RuO_4$ \cite{nakamura13, okazaki13, zhao19a, bertinshaw19, zhang19a,
cirillo19, fursich19}.  Zhao {\it et al.}~have shown that a extremely small
electric current density ($\sim$0.15~A/cm$^2$) can reduce the orthorhombic
distortion and AFM order, and further induce a new orbital
state featuring a simultaneous jump in both magnetization and electrical
resistivity \cite{zhao19a}. The authors propose that orbital occupancies
stabilized by the nonequilibrium current drive the lattice changes and the
novel phenomena.
Because of the distinct energy hierarchy in the SOI system, the narrow band
Mott insulator Sr-214 is more susceptible to such a perturbative approach.  It
is anticipated the crystallographic details through this structural study
would feed critical information for the electronic band structure calculations,
and assist to explain the emerging phenomena including the negative differential
resistivity, reversible resistance switching in Sr-214 \cite{korneta10,
wang15b, cao18}, and the nonlinear conductivity in other spin-orbit-coupled
iridates \cite{cao00,nakano06}.

In summary, neutron diffraction work of the square lattice Sr-214 reveals
a pronounced reduction of the staggered $\rm IrO_6$ distortion below 50 K.
This important observation will help better understand the low-temperature
physics of the archetypal spin-orbit-couple Mott insulator as the magnetic and
transport properties closely tracks the underlying lattice. This point is
further strengthened in that applying a steady electric current at $\rm
8~A/cm^2$ readily reduces the in-plane IrO$_6$ octahedral rotation, which
aligns well with the diminishing bulk magnetization. Our study demonstrates
that the application of electric current is an effective and appealing route
to tune the lattice and probe the rich physics in the SOI system.

We thank Dr. George Jackeli for stimulating discussion. We acknowledge Dr.
Bing Hu for her help in the magnetization measurement and the technical support for
the electric current measurement from Junhong He, Gerald Rucker and Harley
Skorpenske.  This research used resources at the Spallation Neutron Source and
the High Flux Isotope Reactor, which are DOE Office of Science User Facilities
operated by the Oak Ridge National Laboratory. Work at Univ.~of Colorado was
supported by the U.S.~National Science Foundation via grant DMR-1903888.

\end{document}